\documentclass[a4paper,aps,prb]{revtex4}
\usepackage{hyperref}
\usepackage{amsmath,amssymb,mathrsfs}
\usepackage[pdftex]{graphicx}
\usepackage[dvipsnames]{xcolor}
\usepackage[normalem]{ulem}

\begin{document}
\bibliographystyle{unsrt}

\title{Temperature dependent ARPES of the metallic-like bands in Si(553)-Au}

\author{Lenart Dudy}
\affiliation{TEMPO Beamline, Synchrotron SOLEIL, L'Orme des Merisiers Saint-Aubin, B.P.48, 91192 Gif-sur-Yvette, France}
\affiliation{Physikalisches Institut and Wuerzburg-Dresden Cluster of Excellence ct.qmat, Universitaet Wuerzburg, D-97074 Wuerzburg, Germany}

\author{Julian Aulbach}
\affiliation{Physikalisches Institut and Wuerzburg-Dresden Cluster of Excellence ct.qmat, Universitaet Wuerzburg, D-97074 Wuerzburg, Germany}

\author{J\"org Sch\"afer}
\affiliation{Physikalisches Institut and Wuerzburg-Dresden Cluster of Excellence ct.qmat, Universitaet Wuerzburg, D-97074 Wuerzburg, Germany}

\author{Ralph Claessen}
\affiliation{Physikalisches Institut and Wuerzburg-Dresden Cluster of Excellence ct.qmat, Universitaet Wuerzburg, D-97074 Wuerzburg, Germany}

\author{Victor Rogalev}
\affiliation{Physikalisches Institut and Wuerzburg-Dresden Cluster of Excellence ct.qmat, Universitaet Wuerzburg, D-97074 Wuerzburg, Germany}

\affiliation{Diamond Light Source, Didcot, OX11 0DE, UK}

\author{Piotr Chudzinski}
\affiliation{School of Mathematics and Physics, Queen's University of Belfast, UK}
\affiliation{Institute of Fundamental Technological Research, Polish Academy of Sciences, Pawinskiego 5B, 02-105 Warsaw, Poland}

\date{\today}

\begin{abstract}
We conducted a thorough investigation into the temperature dependence of the metallic-like bands of Si(553)-Au using angular-resolved photoemission spectroscopy (ARPES). Our study addresses the challenges posed by the short-term stability of the surface and photo-voltage effects, which we overcame to extract changes in the band-filling and Fermi-velocity. Our findings shed light on the low-temperature phase of the step edge in Si(553)-Au, which has been a topic of ongoing debate regarding its structural or electronic nature. Through comparison with theoretical predictions of a structural-related low-temperature to high-temperature phase transition, we discovered that the band-filling and Fermi-velocity do not change accordingly, thereby ruling out this scenario. Our study contributes to a better understanding of this material system and provides an important reference for future research.
\end{abstract}
\maketitle

\section{Introduction}

\begin{figure}[tbh]
\begin{center}
\includegraphics[width=\linewidth]{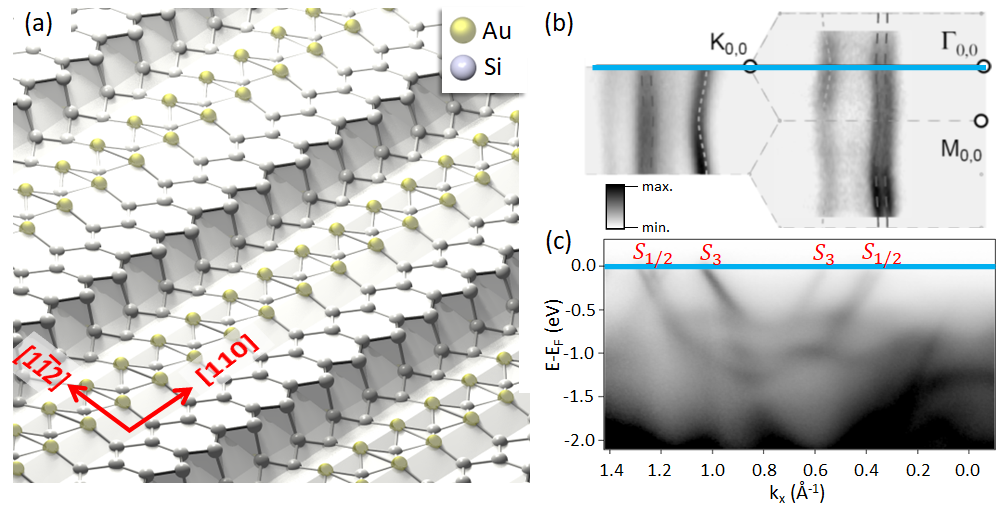}
\caption{(a) Structural model of Si(553)-Au. (b) Fermi-surface at T=50 K with the 1$\times$1 surface Brillouin-zone. (c) ARPES measurement as energy vs. momentum cut showing the Rashba-splitted band S1/S2 and band S3. k$_x$ is along the $[1\bar{1}0]$-direction, along the blue line $\Gamma_{0,0}$-K$_{0,0}$ in (b).  }
\label{Fig:ExplainBands}
\end{center}
\end{figure}

Comprehending a phase transition in the electronic degrees of freedom can be a complex task. One must determine which variable drives the phenomenon and which is driven by it, as well as which variable changes and which does not. Furthermore, experimental research must address the issue of characterizing the pure phenomenon without including unrelated effects. These challenges are particularly relevant to the phase transition of the Si(553)-Au surface reconstruction, which we aim to investigate in the following. The system is a prototype example for a quasi-1D surface system \cite{Snijders2010, Dudy2017}.  A structural model for this surface system was refined by experiment \cite{Ghose2005, Voegeli2010} and density functional theory (DFT) \cite{Erwin2010, Krawiec2010} with details still ongoing \cite{Hafke2016, Braun2020}.  Fig. \ref{Fig:ExplainBands}(a) sketches the structure of this system, consisting of Au-decorated Si(111)-terraces. In a simplified view, there are two electronic subsystems on these terraces: one subsystem is by the structural motive of an Au-double-stranded chain. The other subsystem is located at the zig-zag edge of a Si-honeycomb chain and by 1/3-hole-filled Si-dangling bonds. Both subsystems exhibit distinct periodicities at low temperatures, $\times 2$ for the Au-chain and $\times 3$ for the Si-step edge. The $\times 3$ is only visible at low temperatures in scanning tunneling microscopy\cite{Snijders2012, Song2015} and electron diffraction  \cite{Ahn2005, Edler2019, Hafke2020}. 

Fig. \ref{Fig:ExplainBands}(b) shows the Fermi surface at T=50 K in the 1$\times$1 surface Brillouin-zone measured by ARPES\footnote{The FS and the ARPES measurement of the same measurement was already presented by us in Fig.1 of Ref.  \cite{Chudzinski2021}.}, while Fig. \ref{Fig:ExplainBands}(b) shows the band-dispersion. The metallic-like bands S1 and S2 are a Rashba-split pair \cite{Barke2006}. There is a third metallic band S3. The bands S1 and S2 have \cite{Crain2003, Crain2004, Ahn2005, Chudzinski2021} a filling of about 1/2; band S3 is mostly near 1/3 but varies much stronger due to its higher sensitivity, e.g., to dopants or defects \cite{Song2015}. The orbital origin of the  S1/S2-bands is from hybridization of the Au-atomic chain and the Si-atoms within the Au-ladder and below it (direction in respect to Fig. \ref{Fig:ExplainBands}),  S3 comes from the hybridization of the upper Au-atomic chain with the upper Si-atoms \cite{Krawiec2010, Song2015-2}. The zone-folding at the $\times$2 Brillouin-zone (BZ) at low-T \cite{Ahn2005, Barke2006} was explained in Ref. \cite{Chudzinski2021} as originating from a many-body effect involving a partial-gapping of the bands.

 A key feature of the Si-step edge chain is a $\times 3$ periodicity, long known from STM studies at low temperatures. Studies by SPA-LEED \cite{Edler2019, Hafke2020} found the transition temperature from $\times 3$ to $\times 1$ periodicity to be around 100 K.  Naively, one expects the underlying physics from textbook phenomena like the Peierls-transition built from single-particle states. Then, the expectation is a gap-opening at the Fermi surface. However, while this physics is easy to realize for a strictly 1D system at {\it even} fractional fillings, it is much harder for {\it odd} fillings \footnote{Using standard bosonization techniques, there will be for odd fillings a competition of sine and cosine terms of the same bosonic fields' combination that result in reduced gap size.}. Experiments also suggest that charge density formation by standard Peierls-like mechanisms can be excluded (see Refs. in Ref. \cite{Dudy2017}). Instead, a spin-chain scenario driven by magnetic exchange interactions was proposed \cite{Erwin2010}. Follow-up efforts\cite{Erwin2013} described such phase transition using purely classical methods. Recently, there was the proposal of a structural-induced order-disorder phase transition \cite{Braun2020}. This idea was also advocated by recent experimental results \cite{Mamiyev2021, Plaickner2021}, where Raman spectroscopy detects two variants of local lattice arrangements \cite{Plaickner2021}. In the order-disorder phase-transition model of Ref.~\cite{Braun2020}, a charge transfer from the step-edge to the metallic-like bands is expected. Such change in carrier density of the step-edge can then be discussed to have an effect on the dimerization of the Au-ladder\cite{Mamiyev2018} and, therefore, could act as an explanation alternative to the partial-gapping of the Au-bands advocated by us in Ref.  \cite{Chudzinski2021}.  
 
 This postulated transfer into the metallic bands is about 0.08 electrons per 1$\times$1 surface cell when heating from $\leq$50 K to 200 K (see Fig. 7 in Ref.~\cite{Braun2020}). This sizeable change would be directly visible in temperature-dependent ARPES by a change of band-filling. Also, the model includes a temperature-dependent change of the effective mass (alternatively: Fermi-velocity v$_F$) for the metallic bands. Therefore, it is interesting to experimentally determine how the S1/2 and S3 bands change with temperature. This aim is followed in this work. However, the Si(553)-Au surface system is an experimentally demanding system; it suffers from short-term stability of the surface and the effect of photo-voltage changing with temperature. To succeed, we have to address these issues first. Therefore, the following experimental section \ref{Sec:Experimental} discusses first, how to overcome these issues. Only after that can the experimental extraction succeed, which is presented in the following results section \ref{Sec:Results}. While the inertial data seem to have some temperature dependence for the band-filling, our careful analysis concludes that there is only a very weak change in the band-filling of the metallic-like bands.

\section{Experimental}\label{Sec:Experimental}

\subsection{Preparation}
The preparation was similar to Ref. \cite{Aulbach2017}. We used $n$-doped  Si(553)  substrates (phosphorus-doped, 0.01  $\Omega$ cm). First, the protective photoresist was removed with standard-grade acetone. Then, the samples were cleaned in an ultrasonic bath (2 min. for each solvent) with acetone, isopropanol, and methanol of the highest purity. To avoid residual solvents, the samples were blown off with dry nitrogen. The base pressure of the UHV chamber used for the in situ preparation was below  5$\times$10$^{-11}$ mbar. After degassing the samples,  we heated  up  to 1250$^\circ$C  via  direct  current heating to  remove  the  protective  oxide layer. Gold evaporation of 0.48 ML was performed while the substrate  was  held  at  a  temperature  of  650$^\circ$C,  followed  by a short post-annealing at 850$^\circ$C. The quality of the preparation was controlled by LEED.

\subsection{Surface lifetime and Refreshing procedure}

\begin{figure*}[tbh]
\begin{center}
\includegraphics[width=\linewidth]{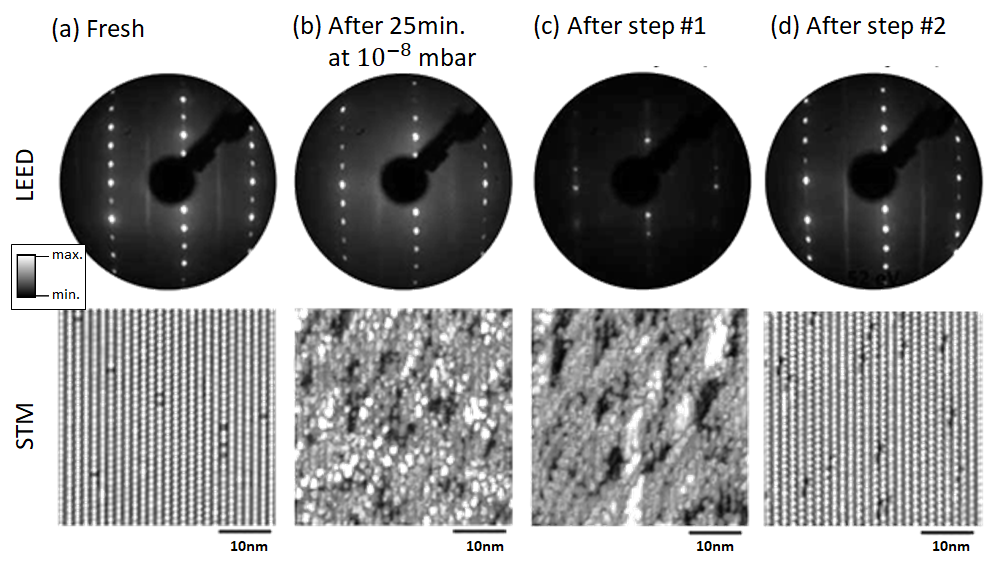}
\caption{LEED patterns (52 eV, T =RT) and STM overview scans (39 nm $\times$ 39 nm, U =+0.5 V, I =5 pA,T =77 K) of a Si(553)-Au sample after each step of the recovery sequence: {\bf (a)} directly after preparation, {\bf(b)} after storage in the load lock at a base pressure of p=5 $\times$10$^{-10}$mbar for 25 min.,{\bf(c)} after the first step of the recovery sequence, {\bf(d)} after the complete sample recovery. No significant differences between the fresh and refreshed samples can be identified.}
\label{Fig:Refresh}.
\end{center}
\end{figure*}

After a particular time\footnote{The time-span is particular because it depends on the method one uses and the quantity one wants to extract as well as pressure, temperature, and details of the UHV system. Anecdotally, we found it occasionally to be only five hours for p=6$\times$10$^{-10}$ mbar and T= 10K when performing ARPES, a full day for p=6$\times$10$^{-11}$ mbar and T= 10K when performing SPA-LEED,  or several days for p=5$\times$10$^{-11}$ mbar and T= 4K when performing STM} of residual gas exposure, a significant loss of sample quality is observed in Si(553)-Au. This degradation was simulated by putting a well-prepared Si(553)-Au sample, see Fig. \ref{Fig:Refresh}(a), into the fast entry load-lock (p=5$\times$10$^{-8}$ mbar) for 25 min. STM and LEED experiments evidence the adsorption of residual gas atoms; see Fig. \ref{Fig:Refresh}(b). The LEED pattern shows an increase in the diffuse background intensity accompanied by a loss of spot intensities. In particular, the $\times$2 streaks are only slightly visible. An STM overview scan discloses a surface completely covered with adsorbates. To recover the sample, the following two-step direct current heating sequence was applied:\\

Step \#1: RT $\xrightarrow[]{t=5s}$ 600$^\circ$C (5s)  $\xrightarrow[]{t=5s}$ RT \\

Step \#2: RT(60s)  $\xrightarrow[]{t=5s}$850$^\circ$C (5s)   $\xrightarrow[]{t=5s}$ RT \\

The first step aims to desorb residual gas atoms and is accompanied by a short pressure boost. The second step ensures a reassembly of the surface after adsorbate desorption. During the second step, no further pressure boost occurs. The sample refresh is analyzed by LEED and STM experiments after step \#1 and step \#2 of the recovery sequence; see Fig. \ref{Fig:Refresh} (c) and (d). After step \#1, no Si(553)-Au spots but rather (1$\times$1) spots are observed. The STM overview scan reveals a rough, disordered sample surface. After step \#2, LEED and STM images indicate a well-ordered Si(553)-Au atomic wire array. No significant difference is visible when comparing the LEED image to the freshly prepared sample; cf. Fig. \ref{Fig:Refresh} (a) and (d). There is only a slight increase in the number of defects visible in the STM image. 

This process can be repeated many times without significantly losing sample quality visible in LEED and STM. Later on, we will see only a small irreversible component in our ARPES result. Nevertheless, the quality is very high after many repetitions. This contrasts the one-step process suggested in Refs. \cite{Altmann2001} and  \cite{Okuda2010}, which can only be repeated a limited number without sample degradation. A plausible explanation for this difference may be a considerable loss of Au-atoms when combining adsorbate desorption and surface reordering in one step. 

\subsection{Temperature-dependent ARPES: Surface Quality and Surface Photovoltage}

\begin{figure*}[bth]
\begin{center}
\includegraphics[width=\linewidth]{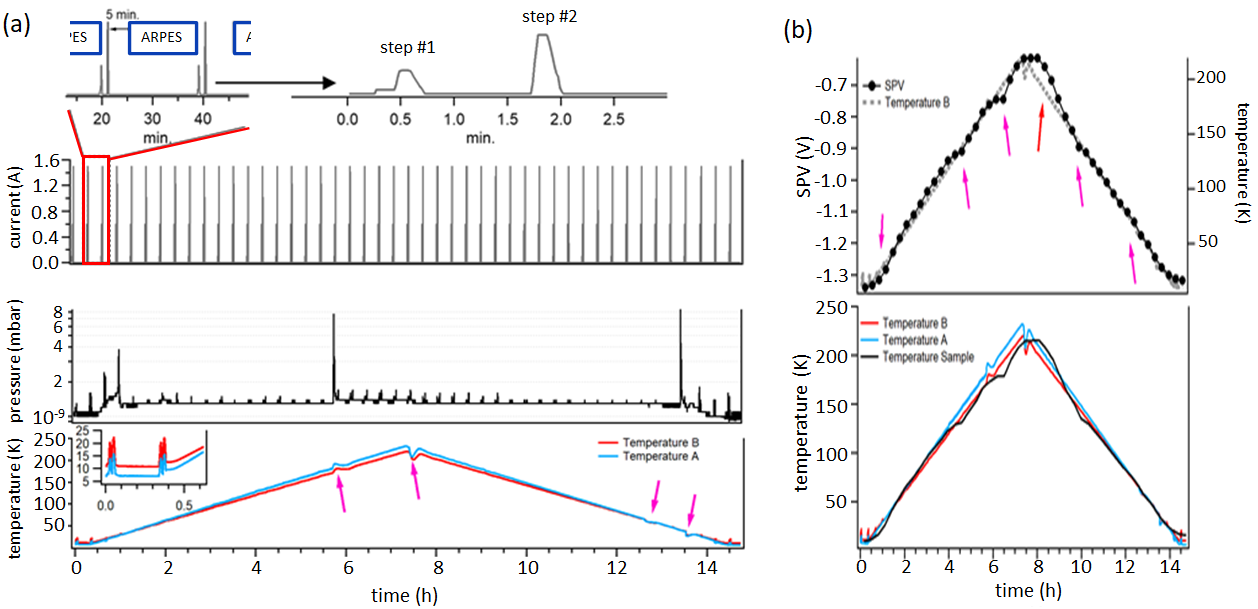}
\caption{(a) The entire ARPES measurement cycle for some selected variables of interest. The upper panel shows the power supply's current, performing the sample's DC heating with the refresh procedure described above. The current was adjusted to the temperatures by a pyrometer. The middle panel shows the pressure evolution during the experiment. The pressure was 6$\times$10$^{-10}$ mbar at low temperatures. The cryostat temperature was ramped at a constant rate and controlled by the liquid helium flux and a resistive heater. The pink arrows mark the position where the helium flux was changed manually. (b) The upper panel shows the change in surface photo-voltages (SPV) during the measurements, following overall linearly the temperature change (temperature B with the scale on the right). The lower panel shows the sample temperature as extracted from the SPV and compared to the readings of the temperature sensors.}
\label{Fig:ARPES_T_proc}
\end{center}
\end{figure*}

\begin{figure}[hbt]
\begin{center}
\includegraphics[width=0.8\linewidth]{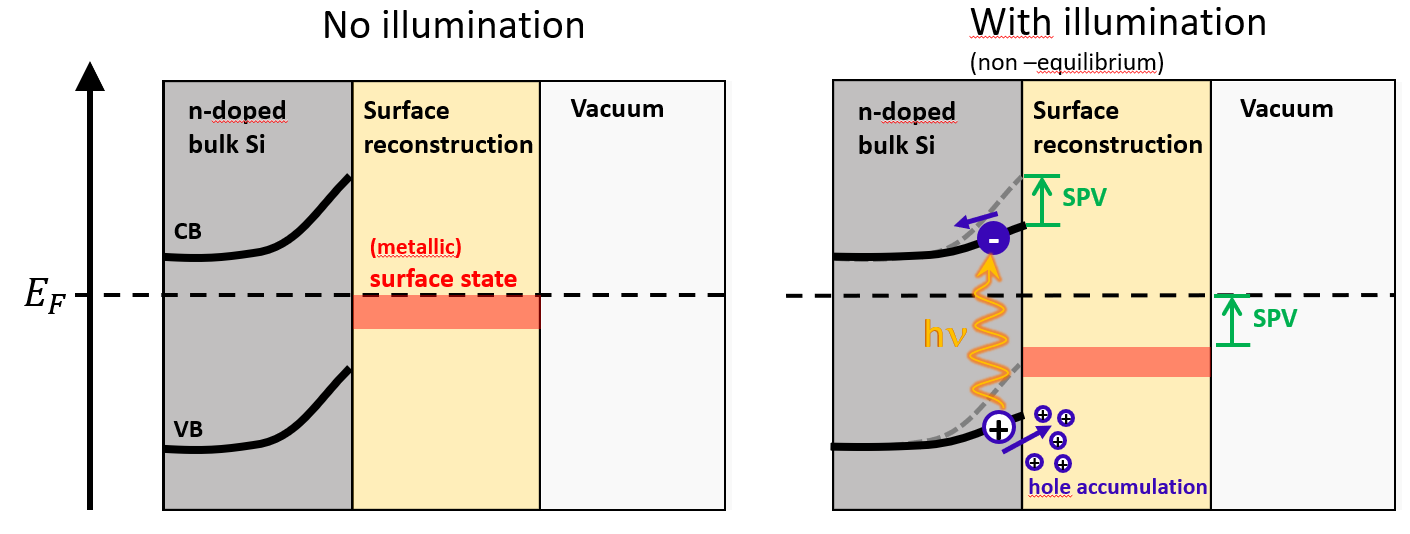}
\caption{Sketch to explain the surface photo-voltage (SPV): (Left) The n-doped Si shows band-bending  near the metallic overlayer of the surface reconstruction (forming a Schottky barrier). (Right) With illumination, electrons and holes are created. At low temperatures, the reduced surface recombination causes accumulated charge. A non-equilibrium electric field  appears which reduces the band bending (the resulting potential difference is labeled SPV).}
\label{Fig:ExplainSPV}
\end{center}
\end{figure}

Photoemission measurements were performed with a non-monochromatized duo-plasmatron He-discharge lamp. The base pressure of the analysis chamber was $\approx$7$\times$10$^{-11}$ mbar. The operational pressure in the analysis chamber was p$\approx$6$\times$10$^{-10}$ mbar to 1$\times$10$^{-9}$ mbar (mainly due to the Helium partial pressure but also due to evaporates from the cryostat depending on the temperature, see pressure in Fig. \ref{Fig:ARPES_T_proc} (a) middle panel).  We used the He-I line at h$\nu$ = 21.22 eV. The energy resolution of the spectrometer was 25 meV. Despite a residual gas filter for the He-gas line, the sample is prone to degradation under the pressure of the He-lamp.  Therefore, we performed a thermal refresh procedure as described above before each ARPES measurement.

The measurement procedure is illustrated in Fig. \ref{Fig:ARPES_T_proc} (a). Before the measurement, the sample was thermalized for about 1h at 10 K. See the lower panel of Fig. \ref{Fig:ARPES_T_proc} (a). Sensor A is a silicon diode near the cooling head, and sensor B is about 10 mm away from the sample. The refresh described above was applied after each measurement step. The upper panel of Fig. \ref{Fig:ARPES_T_proc} (a) shows the read-back of the current source doing the refresh by direct current heating. In the middle panel of Fig. \ref{Fig:ARPES_T_proc} (a), one also sees a slight rise in pressure, especially for low T, when the refresh is done. After the direct current heating sequence, the temperature reached the base temperature after around 2 mins, as inferred from the low temperatures (see Fig. \ref{Fig:ARPES_T_proc} (a), inset of the lower panel). The sample was then given 5 min. for reaching equilibrium before starting the ARPES acquisition at the read temperature. For analysis, the normalization is by the photon dose using the same current of the He-discharge lamp and the same counting time.

Si(553)-Au shows a rather strong surface photo-voltage (SPV). The principle of SPV is sketched in Fig. \ref{Fig:ExplainSPV} and briefly explained as follows (for more details, see Refs. \cite{Kronik1999} and \cite{Hecht1990}, more specific examples for photoemission spectroscopy can be found, i.e., in Refs.\cite{Demuth1986} and \cite{Alonso1990} ). To the left of Fig. \ref{Fig:ExplainSPV}, we see the situation of an n-doped Si forming a Schottky barrier in the presence of the metallic overlayer of the surface reconstruction. With illumination, see right side of Fig. \ref{Fig:ExplainSPV}, electrons and holes are created. At room temperature, this charge recombines sufficiently fast not to alter the equilibrium electrostatic potential in the surface region. At low temperatures, the reduced surface recombination allows the accumulated charge at the semiconductor-overlayer interface to set up a non-equilibrium electric field (the SPV) which reduces (or even cancels) the band bending.

 During the measurements, we covered  all viewports to prevent light pollution and performed the measurements under a stable photocurrent of the He-discharge lamp. Because of the SPV changing with temperature, we have to align the photo-electron kinetic energies by a line-fit of a Fermi function. The zero reference for the SPV given in Fig.\ref{Fig:ARPES_T_proc} (b) is the Fermi-energy of an Au(111) sample. The obtained SPV shows clearly a linear relation with temperature. Compared to a Schottky barrier-like situation (cf. Ref. \cite{Hecht1990}, see there especially Fig. 2), the visible T-linearity is likely caused by the high  doping combined with the intense  lamp. This is also obvious from the relatively large magnitude of the shift in SPV.  
 As a detail, the red and pink arrows in Fig. \ref{Fig:ARPES_T_proc} (b) coincide with substantial changes either in the helium flux or resistive heating. In order to check the temperature scan, we can use the sample itself, as a built-in Schottky-diode, for temperature calibration. We assume that temperature B represents at start condition (after long thermalization) the sample temperature and that the curves of temperature B and sample temperature coincide on average, and have, therefore, the smallest $\chi^2$. These both assumptions result in the formula
\begin{eqnarray*}
SPV(T)&=&(-1.3734\,eV \pm 6.3\,meV)  \\
 &+& (3.5\,meV/K \pm 0.05\,meV/K)\times T 
\end{eqnarray*}

The inverse function, T(SPV), gives the sample temperature displayed in the lower panel of Fig. \ref{Fig:ARPES_T_proc} (b). Note that the temperature correction does not strongly alternate the final results. However, the dependence of the SPV, as a direct probe of the samples’ temperature, proves that we followed a very systematic heating/cooling cycle. We will soon discuss contributions in this cycle which are reversible and which are irreversible.

\begin{figure}[hbt]
\begin{center}
\includegraphics[width=0.5\linewidth]{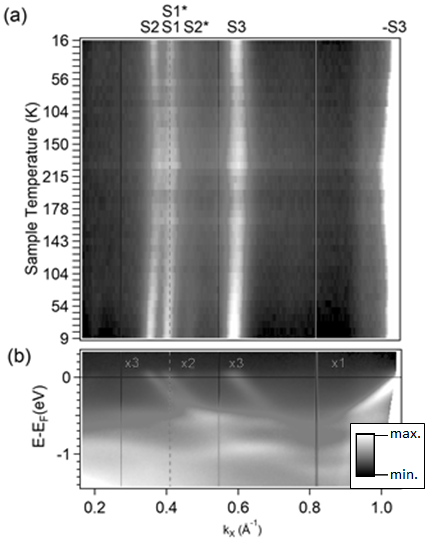}
\caption{(a) The temperature dependence of the MDC at E$_F$. (b) ARPES map at T=9 K as an aid for understanding the band topology as well as the labels on the top. A small shift of the Fermi-wavelength k$_F$  is visible in the MDCs of (a), showing a reduced number of carriers at high T.}
\label{Fig:ARPES_T_MDC}
\end{center}
\end{figure}

\section{Results} \label{Sec:Results}

We will yet discuss the ARPES spectra. For each temperature, the spectrum was measured as $I(E,k)$, the intensity of photoelectrons in dependence on energy and momentum. An example of such spectrum, here called ARPES map, is shown in Fig. \ref{Fig:ARPES_T_MDC}. As for the momentum in the surface Brillouin zone, we measured along the blue line given in Fig. \ref{Fig:ExplainBands} (a).

\subsection{Change of Fermi-wavelength and band filling with temperature}

\begin{figure*}[tbh]
\begin{center}
\includegraphics[width=\linewidth]{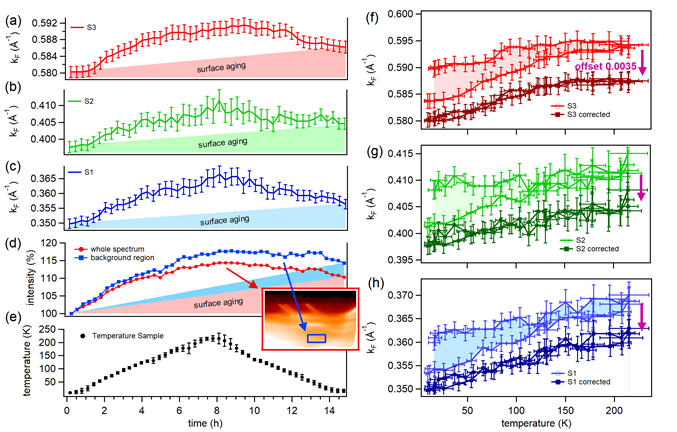}
\caption{(a)-(d) Results of the line-fit of the MDC’s shown in Fig. \ref{Fig:ARPES_T_MDC}. Displayed are the extracted Fermi-wavelengths for bands S1-S3. The changes in k$_F$ are considered to have an irreversible (surface aging) and a reversible component. (d) A similar two-component view applies to the photoemission intensity, either integrated over the full range or in a background region. The irreversible component can be also seen in (f)-(h) where the results of (a)-(c) are plotted against (e) the sample temperature. Please note that in (f)-(h) the uncorrected S1-S3 are shifted up by 0.035 \AA$^{-1}$ (see purple arrow) to improve the visibility of the corrected k$_F$ (T). The corrected k$_F$ (T) for the bands S1-S3 is achieved by assuming the irreversible component to be linear and subtracting it. }
\label{Fig:ARPES_T_MDC_fit}
\end{center}
\end{figure*}

During the temperature scan, we observed a small change in the Fermi-wavelengths (k$_F$'s) of the metallic bands which we will discuss in the following. This change can be observed by analyzing the so-called momentum distribution curves (MDCs) at the Fermi-energy (E$_F$). To obtain MDCs, one slices the experimental ARPES map shown in Figure \ref{Fig:ARPES_T_MDC}(b) at fixed energy (here E$_F$) along the momentum direction. The maxima in the MDCs at E$_F$ are the Fermi-momenta of the corresponding bands.

Figure \ref{Fig:ARPES_T_MDC}(a) illustrates the temperature dependence of the MDCs at E$_F$. These MDCs were integrated within a $\pm$ 25 meV range. The MDCs from the temperature cycle are arranged vertically in the plot, with the MDCs from low temperatures (start of cycle) at the bottom, the MDCs from the highest temperature in the middle, and the MDCs from low temperatures (end of cycle) again at the top. To better understand the band topology and the MDCs, refer to compare Figure \ref{Fig:ARPES_T_MDC}(a) and (b). 

In Fig. \ref{Fig:ARPES_T_MDC}  (a), all the metallic bands move towards lower filling (the electron-like parabolas of Fig. \ref{Fig:ARPES_T_MDC}  (b) get smaller) when starting from the bottom at low T and going to higher T in the middle. This bend reverses after reaching the middle with the cool-down. The effect is not large (below 1\%) but visible. Compared to the situation expected by the theory of Ref. \cite{Braun2020}, where the charge-transfer causes the number of electrons to increase with temperature for the subsystem of the metallic bands, note that, actually, with increasing temperature, the number of electrons is reduced.

\subsubsection{Irreversible and reversible components}

In order to extract the electron count, we determine the Fermi-wavelength k$_F$ and then use the Luttinger theorem, allowing to calculate the electrons in the band by the Fermi momentum, i.e., a half electron filled band occupies half of its Brillouin zone. For obtaining the k$_F$'s, we performed a line-fit of the MDC’s at E$_F$ using three Lorentzians (ignoring contributions of the replica-bands S1*, S2*). The changes in Fermi-wavelength k$_F$   are shown in Fig. \ref{Fig:ARPES_T_MDC_fit} (a)–(c). Instead of being completely reversible, we notice an asymmetry indicating an irreversible component. We, therefore, consider the changes consisting of two components: One component is irreversible, and the other reversible. The irreversible component is considered to stem from the aging/deterioration of the surface.

We can probe the consideration that the irreversible component results from surface deterioration; a deteriorated surface enhances the inelastic scattering of outgoing electrons. Such inelastic scattering will be strongly visible by stronger background. In Figure \ref{Fig:ARPES_T_MDC_fit} (d), we present a comparison of two regions. The inset illustrates the specific regions where the intensity was measured, with red representing the entire ARPES map and blue indicating a selected background region. The comparison reveals that (i) in both regions the intensity changes are comparable (showing that a change in background is the main effect also visible for the entire ARPES map) and (ii) the background intensity is higher at the end of the cycle.  Therefore, our assumed scenario, where an irreversible component is attributed to sample deterioration, aligns with reality.

In all the situations mentioned earlier, both the Fermi momentum and the background intensity exhibit a consistent linear increase as the measurement time progresses, indicating an irreversible aging effect on the surface. To address this systematic error, we can apply a correction by subtracting the linear-T component. This correction restores the reversible behavior, as shown in Figure \ref{Fig:ARPES_T_MDC_fit} (f)-(h). Prior to the correction (light colors), the heating and cooling phases of the cycle do not exhibit the same Fermi momentum (k$_F$) at equivalent temperatures of the cycle (as indicated by the shaded area). However, after the correction (dark colors), this discrepancy is resolved.

It's important to note that this correction is not directly dependent on temperature, but rather on time. It assumes that a similar amount of defects is introduced with each refresh, contributing to the irreversible component. However, for this correction to be effective, it is crucial that the temperatures at the beginning of the cycle (T=9 K) and the end of the cycle (T=16 K) are nearly equal. If they are significantly different, we could not estimate the irreversible component.

However, after applying this correction, we finally now turn to the reversible behavior of the band-filling. The reversible behavior is mainly a linear T-dependent change in the filling of the metallic bands; see corrected curved in Fig. \ref{Fig:ARPES_T_MDC_fit} (f)-(h). The k$_F$ of band S3 deviates from the linear T-dependence for high temperatures. It shows saturation. We can yet conclude with the change in the filling. As stated before the filling can be calculated by the Fermi-momenta using the Luttinger theorem. By considering all the metallic bands and comparing the lowest and highest temperature, our data displays a loss for the metallic subsystem of about 0.03$\pm$0.01 electrons per 1$\times$1 surface cell. 

\subsection{Fermi-velocity with temperature}
\begin{figure}[hbt]
\begin{center}
\includegraphics[width=0.45\linewidth]{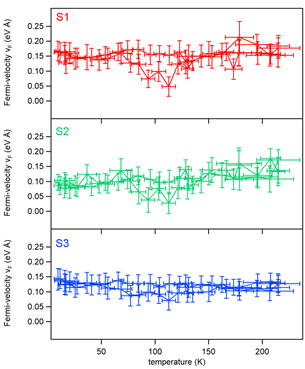}
\caption{Fermi-velocity of the three bands vs. temperature. For details see text.}
\label{Fig:ARPES_T_vF_fit}
\end{center}
\end{figure}

We determined the Fermi-velocity as follows: The dispersion of each band was obtained by the maximum position of the MDC’s in the range of E-E$_F$=-0.1 eV to 0 eV. For this, each MDC was fitted with three Lorentzian for the bands S1-S3 (ignoring the replica bands S1*/S2*). After that, the Fermi-velocity v$_F$ (more precisely: the Fermi-velocity projected along k$_x$) could be extracted by a line-fit with the function E(k)-E$_F$=v$_F$ (k-k$_F$). The results of this analysis are displayed in Fig. \ref{Fig:ARPES_T_vF_fit}. The average values are v$_F$=(0.15 $\pm$ 0.05) eV\AA\ ($\approx 0.23\times 10^5$ m/s) for S1, v$_F$=(0.12 $\pm$ 0.06) eV\AA\ ($\approx 0.18\times 10^5$ m/s) for S2, and v$_F$=(0.12 $\pm$ 0.05) eV\AA\ ($\approx 0.18\times 10^5$ m/s) for S3. There are slight changes in S2 (and maybe some systematic derivations of S1/S3 around 100 K) which we attribute to the systematic error of our fit-model neglecting the replica bands. Nevertheless, the key result here is that the Fermi-velocity of each metallic stays almost constant with temperature. 

\section{Discussion}

We see that the SPV has a large effect on the shift of the ARPES spectra. Therefore, we advise caution when interpreting low-temperature gap values obtained by ARPES, such as those reported in Ref. \cite{Ahn2005}, as changes in light intensity can alter the SPV  and ARPES spectra. For this, it is interesting to compare the magnitude of the photo-voltage (being around 1 eV) with theoretically expected gaps (a few meV's).  In seeing this, we think that more work has to be done to measure real gap values of the metallic-like bands by ARPES. Nevertheless, the existence of solitons in the Au-double chain \cite{Chudzinski2021} proves experimentally the existence of a (partial) gap which is theoretically accessed \cite{Chudzinski2021} to be of the order of 14 meV.

Our investigation showed that the filling of the metallic-like band subsystem decreases by approximately 0.03 electrons per 1$\times$1 surface cell when the temperature increases from $\sim$10 K to 260 K.  The change is mainly linear for S1/2, there is a slight non-linearity in S3 at high temperatures. Because of the T-linear changing surface photovoltage, we speculate that the surface photovoltage generates the experimentally observed T-linear changes in band-filling.Presumably, in the presence of the strong SPV, the photo-generated holes cannot be completely replenished due to limited surface conductivity.

We compared our findings to the phase-transition model proposed in Ref. \cite{Braun2020}. In agreement with SPA-LEED results\cite{Edler2019, Hafke2020}, we can assume that the phase transition of the $\times 3$ edge state is completely covered within the range of temperatures that we have probed. In the model of Ref. \cite{Braun2020},  we would therefore probe the full possible charge transfer. However, our results do not agree with this: first of all, the charge transfer with temperature is opposite. In theory, the subsystem of the metallic bands is expected to {\it gain} about 0.08 electrons when heating from $\leq$50 K to 200 K, in the experiment, 0.03$\pm$0.01 electrons are {\it lost}. In theory, the effective masses show a significant change, including a sign change of the effective electron masses. In the experiment, the Fermi-velocities (and therefore the effective electron masses) stay almost constant.

\section{Conclusions}

In summary, our study employed ARPES to investigate the temperature dependence of the metallic bands in Si(553)-Au. To address the issue of adsorbates, we optimized a sample refreshing procedure. We then addressed the impact of surface voltage and its variation with temperature. Subsequently, we discovered that changes in the band filling and Fermi velocity of the metallic-like bands were relatively minor across the temperature range studied. Our results do not mean that, generally, all models considering an order-disorder crossover are ruled out. However, our findings do not agree with any model requiring a charge transfer from the metallic bands to the step edge changing strongly with temperature. Our results raise the question of whether ARPES is useful to detect any changes during the phase transition of the Si-edge in Si(553)-Au. Our results show that with careful data analysis such information can indeed be extracted with remarkable precision, leading in our case to the finding that there is (almost) no change for the Fermi wavelength and Fermi velocity of the metallic bands. However, in forthcoming publications\cite{Chudzinski2022, Chudzinski2023}, we will discuss the strong sensitivity of ARPES for the phase transition of the Si-edge showing the temperature-dependence of the long-missing S4 band excitation of the step-edge. 

\bibliographystyle{elsarticle-num-names} 
\bibliography{AuSi553ARPES}





\end{document}